\documentclass[12pt,letterpaper]{article}
\usepackage[a4paper, total={7in, 10in}]{geometry}

\usepackage{graphicx}
\usepackage{helvet}
\usepackage{authblk}
\usepackage{hyperref}
\usepackage{amsmath} 
\usepackage{amssymb}
\usepackage{etoolbox}  
\usepackage{orcidlink} 
\usepackage[super,comma,sort&compress]  
   {natbib}

\newcommand{\be}{\begin{eqnarray}}
\newcommand{\ee}{\end{eqnarray}}

\makeatletter
\renewcommand{\maketitle}{\bgroup\setlength{\parindent}{0pt}
\begin{flushleft}
  \textbf{\@title}
  
  \@author
\end{flushleft}\egroup}
\makeatother

\title{An interstellar mission to test astrophysical black holes}
\date{}

\author[1,2,*,\orcidlink{0000-0002-3180-9502}]{Cosimo Bambi}

\affil[1]{Center for Astronomy and Astrophysics, Center for Field Theory and Particle Physics, and Department of Physics,
Fudan University, Shanghai 200438, China}
\affil[2]{School of Natural Sciences and Humanities, New Uzbekistan University, Tashkent 100007, Uzbekistan}

\affil[*]{Correspondence: bambi@fudan.edu.cn}

\begin{document}

\maketitle

\section*{SUMMARY}

Black holes are the sources of the strongest gravitational fields that can be found today in the Universe and are ideal laboratories for testing Einstein's theory of General Relativity in the strong field regime. In this letter, I show that the possibility of an interstellar mission to send a small spacecraft to the nearest black hole, although very speculative and extremely challenging, is not completely unrealistic. Certainly we do not have the necessary technology today, but it may be available in the next 20-30~years. The mission may last 80-100~years, but we would be able to obtain very valuable information about black holes and General Relativity that might be difficult to obtain in other ways.

\section*{KEYWORDS}


Black holes, Interstellar missions, Tests of General Relativity, Nanocrafts, Lightsails, Laser propulsion

\section*{INTRODUCTION}

Einstein's theory of General Relativity is our current framework for the description of gravitational interactions and the spacetime structure. For decades, the theory has been extensively tested in the so-called weak field limit with experiments in the Solar System and radio observations of binary pulsars~\cite{Will:2014kxa}. The strong field regime was almost completely unexplored up to 10~years ago.

Black holes are ideal laboratories for testing Einstein's theory of General Relativity in the strong field regime because they are the sources of the strongest gravitational fields that we can found today in the Universe~\cite{Bambi:2015kza,Bambi:2017khi}. Only 10~years ago, we knew stellar-mass black holes as compact objects in X-ray binary systems with masses exceeding the maximum mass for a neutron star~\cite{Rhoades:1974fn} and supermassive black holes in galactic nuclei as systems that were too heavy and too compact to be clusters of non-luminous bodies like neutron stars~\cite{Maoz:1997yd}. There was no evidence that the spacetime geometry around these objects was described by the Kerr solution as predicted by General Relativity~\cite{Kerr:1963ud}.

The past 10~years have dramatically changed this research field and today we can test the Kerr metric around black holes -- and, more in general, we can use black holes for testing fundamental physics -- with gravitational waves~\cite{LIGOScientific:2016lio,LIGOScientific:2019fpa}, X-rays~\cite{Cao:2017kdq,Tripathi:2018lhx,Tripathi:2020yts}, and black hole imaging~\cite{EventHorizonTelescope:2020qrl,Vagnozzi:2022moj}. All available observations are consistent with the predictions of General Relativity and current constraints can be improved by the next generation of observational facilities. However, the constraining capability with these techniques will be likely limited by complications of the astrophysical environment and simplifications in our theoretical models. In general, it will be very difficult to get very precise and accurate measurements.


\section*{THE CLOSES BLACK HOLE TO EARTH}

As of now, the closest {\it known} black hole to Earth is GAIA-BH1, which was discovered in September~2022 and is at a distance of 478~pc (1,560~light-years)~\cite{El-Badry:2022zih,Chakrabarti:2022eyq}. However, we can expect that there are many {\it unknown} black holes closer to Earth. From simple considerations, we can estimate that the closest black hole to Earth may be at only 20-25~light-years, even if this is just a rough estimate and is affected by large uncertainties.

We estimate that in the Milky Way there are between $1 \cdot 10^{11}$ and $4 \cdot 10^{11}$ normal starts. Current models suggest that there are about $1 \cdot 10^{10}$~white dwarfs in the Milky Way, but a significant fraction may be in the Galactic halo and only $2 \cdot 10^9$~white dwarfs may be in the Galactic thin disk~\cite{Napiwotzki:2009tf}. Assuming that the maximum mass for a neutron star is 1.7~$M_\odot$, Timmes et al.~(1996)~\cite{Timmes:1995kp} predicted $1.4 \cdot 10^9$~stellar-mass black holes in the Milky Way. Today we know that the maximum mass for a neutron star is somewhat higher~\cite{Antoniadis:2013pzd}, but still we can expect to have around $1 \cdot 10^9$~stellar-mass black holes in the Galaxy. While there are large uncertainties on all these numbers, we can roughly estimate that in the Milky Way there are 1~black hole and 10~white dwarfs every 100~normal stars. Their distributions are not perfectly homogeneous, but here we only want to get a rough estimate of the possible distance of the closest black hole.

\begin{figure}
\centering
\includegraphics[width=1\linewidth]{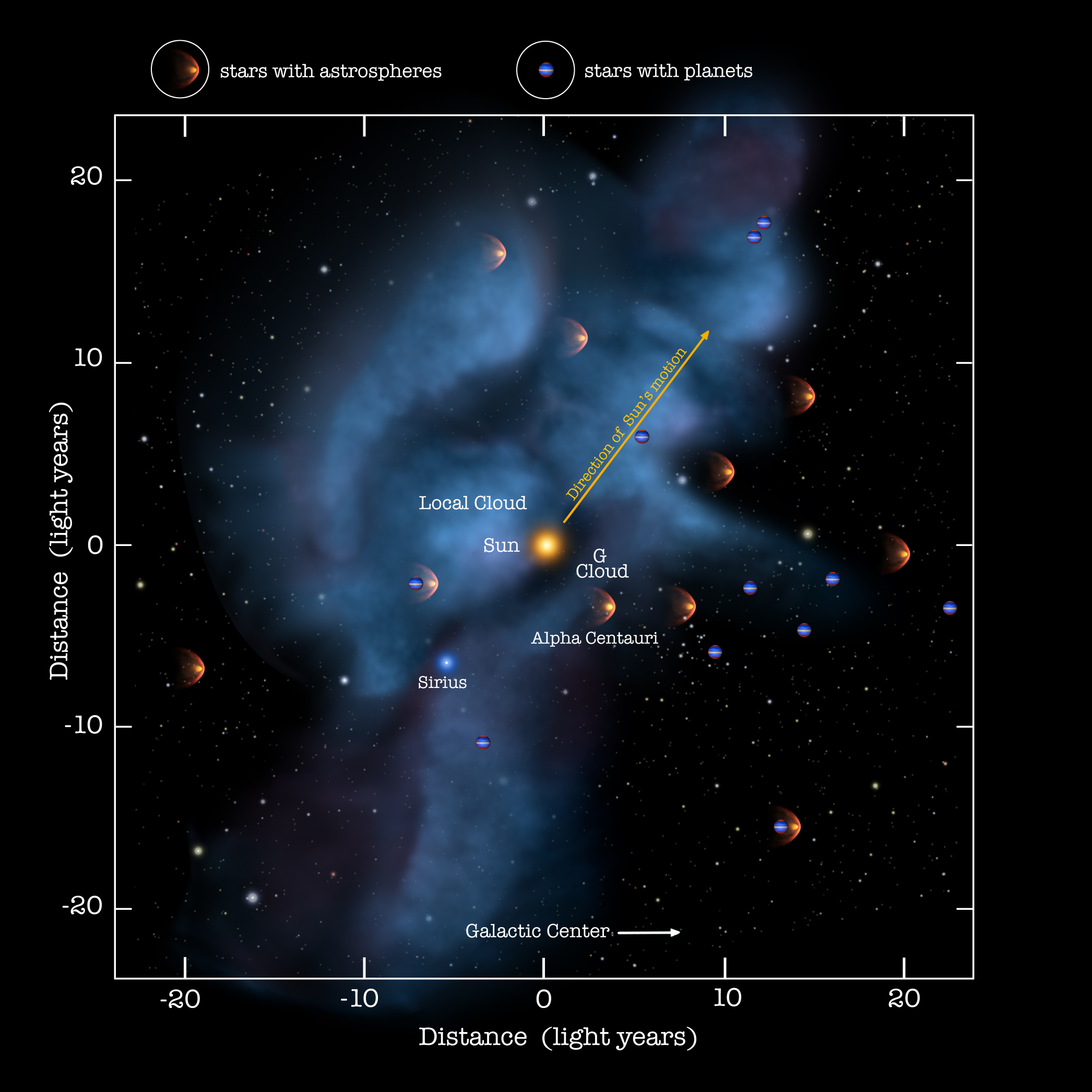}
\caption{{\bf Stars and exoplanets within 25~light-years of the Earth.}
It is plausible that in this region there is (somewhere) even a black hole. Credit: NASA Goddard/Adler/U. Chicago/Wesleyan.
\label{f-local}}
\end{figure}

Within 5.0~pc (16.3~light-years) of Earth, we know about 60~main sequence stars (most of them are red dwarfs, like Proxima Centauri, which is at a distance of 4.2~light-years and is the closest star to us) and 4~white dwarfs (Sirius~B, Procyon~B, van Maanen's Star, and Gliese~440)~\cite{Perryman:1997sa}. Within 20.0~pc (65.2~light-years) of us, we know about 130~white dwarfs, but they may be more, maybe around 160~white dwarfs in total~\cite{Sion:2009ur}. From the previous rough estimate (1~black hole and 10~white dwarfs every 100~normal stars), we can argue that the closest black hole may be within 6-8~pc of Earth, roughly corresponding to 20-25~light-years (see Fig.~\ref{f-local}).

If our closest black hole is in a binary system with a normal star, it could be discovered by studying the orbital motion of the companion star. If it is in a binary system with a neutron star or another black hole, or it is an isolated black hole, its detection is definitively more challenging. Olejak et al.~(2020)~\cite{Olejak:2019pln} estimated that more than 90\% of the black holes in the Galactic disk are isolated objects, with no companion. Currently, the only technique to detect an isolated black hole is through microlensing and so far only a black hole has been clearly discovered in this way, OGLE-2011-BLG-0462, which is at distance of 1.6~kpc~\cite{OGLE:2022gdj}. In the future, we may be able to use novel techniques to detect isolated black holes. Recently there have been some proposals to discover nearby isolated black holes.

Murchikova \& Sahu (2025)~\cite{Murchikova:2025oio} proposed to use observational facilities like the Square Kilometer Array (SKA), the Atacama Large Millimiter/Submillimiter Array (ALMA), and James Webb Space Telescope (JWST). Isolated black holes moving through the interstellar medium can accrete from the interstellar medium itself and such an accretion process produces electromagnetic radiation. Murchikova \& Sahu (2025)~\cite{Murchikova:2025oio} showed that current observational facilities can already detect the radiation from isolated black holes in the warm medium of the Local Interstellar Cloud within 50~pc of Earth, but their identification as accreting black holes is challenging and requires multi-telescope observations.

\begin{figure*}[t]
\centering
\includegraphics[width=0.27\linewidth]{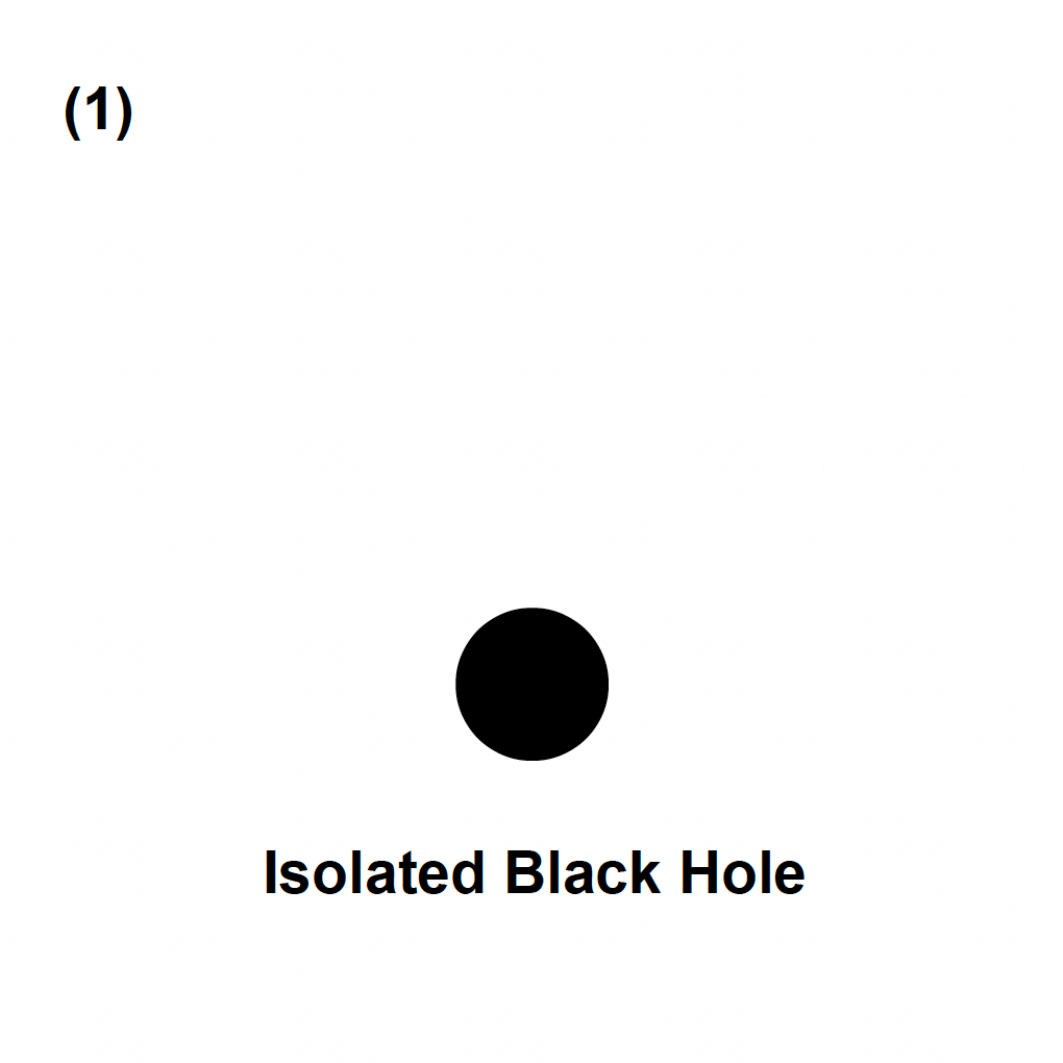}
\hspace{1.0cm}
\includegraphics[width=0.27\linewidth]{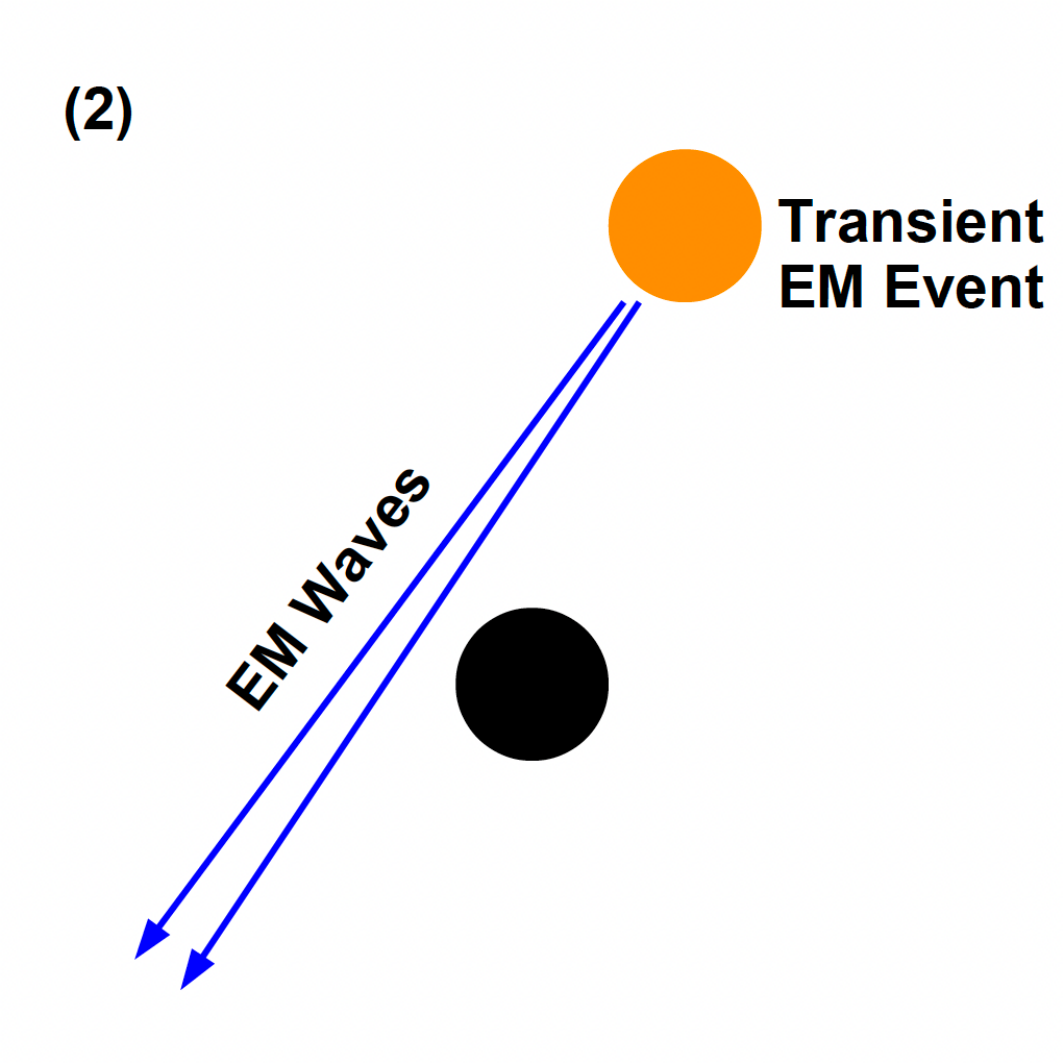}
\hspace{1.0cm}
\includegraphics[width=0.27\linewidth]{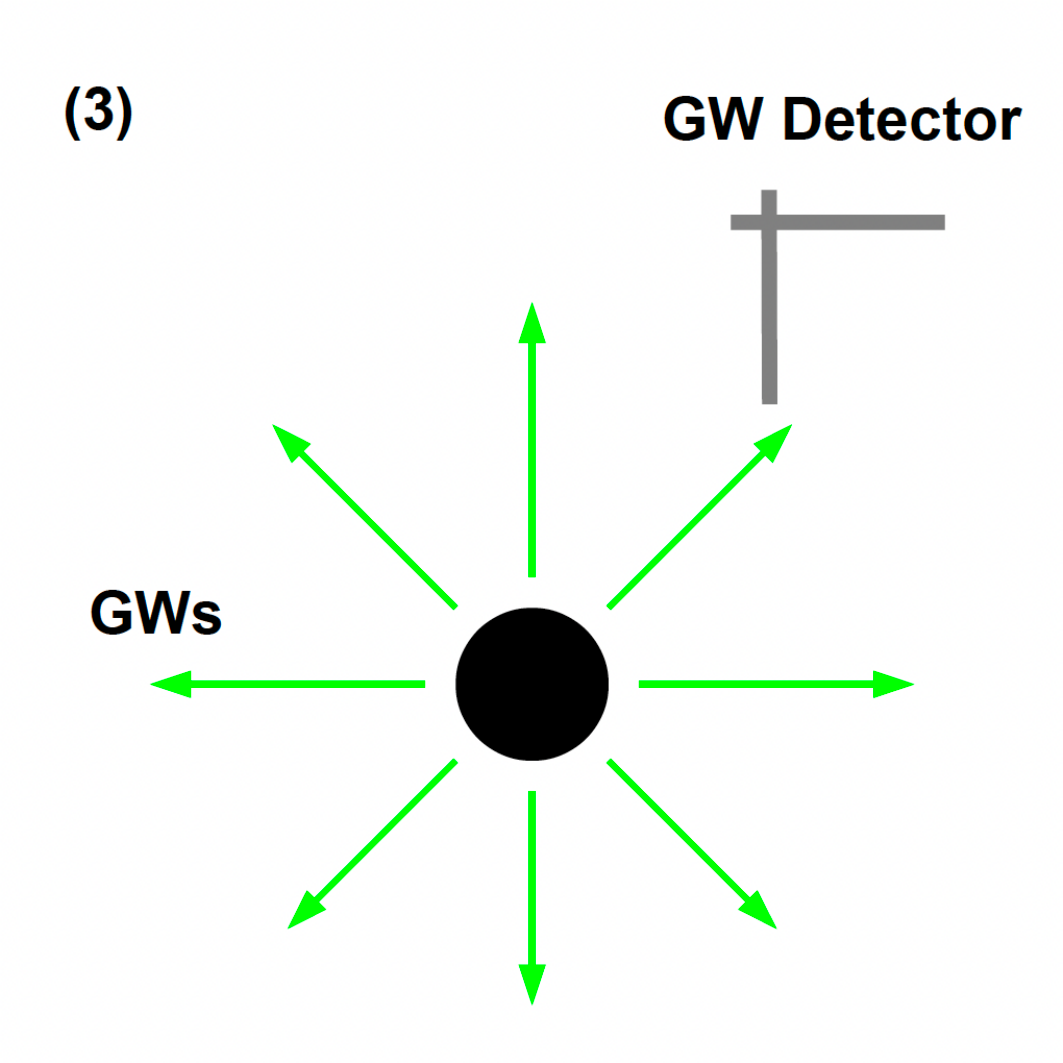}
\caption{{\bf Detection of a nearby isolated black hole as proposed in Jana et al.~(2025)~\cite{Jana:2024hks}.} $(1)$ An isolated black hole is not far from Earth. $(2)$ A transient electromagnetic event illuminates the isolated black hole and distorts the spacetime around the black hole. $(3)$ The spacetime emits gravitational waves to return to its equilibrium configuration and these gravitational waves can be detected by our ground-based laser interferometers.
\label{f-GW}}
\end{figure*}

Jana et al.~(2025)~\cite{Jana:2024hks} proposed to detect isolated black holes with gravitational waves: transient electromagnetic events could excite the spacetime around an isolated black hole, which should thus emit gravitational waves (see Fig.~\ref{f-GW}). They estimated that LIGO~A+ should be able to detect isolated black holes within 50~pc of Earth with this technique~\cite{Jana:2024hks}.

\section*{INTERSTELLAR MISSION}

In what follows, we want to discuss the possibility of sending a spacecraft to a black hole at 20-25~light-years from Earth to test the nature of the compact object and fundamental physics in strong gravitational fields. Clearly the spacecraft should be able to travel at some fractions of the speed of light, which would be certainly impossible with chemically propelled spacecrafts. The Tsiolkovsky rocket equation reads $m_i/m_f = e^{\Delta v / v_e}$, where $m_i$ is the initial total mass of the rocket (including propellant), $m_f$ is the final total mass of the rocket (without propellant), $\Delta v$ is the total change of the rocket's velocity, and $v_e$ is the effective exhaust velocity. For example, the American Space Shuttle and the European Ariane launchers use liquid hydrogen/liquid oxygen and $v_e \sim 4.5$~km/s. If we want the rocket to travel at 1/10~of the speed of light, $m_i/m_f \sim 10^{28953}$: even if $m_f$ were the proton mass, $m_i$ would exceed largely the total mass of the visible Universe. Gravity-assist techniques~\cite{LPS17,Vasile06,Pan23} commonly used to increase spacecraft velocities cannot certainly solve this problem.

As of now, {\it nanocrafts}~\cite{Lubin16,Lubin22,Kuhlmey25} seem to be the most promising solution for interstellar missions, even if we do not have yet the technology. A nanocraft is a gram-scale spacecraft comprising two main parts: a gram-scale wafer and a light sail. The gram-scale wafer is the main body of the satellite and constitutes a fully functional space probe: it has a computer processor, thrusters, solar panels, navigation and communication equipment, etc. The light sail is an extremely thin, meter-scale, dielectric metamaterial used to accelerate the probe. The idea is to use ground-based high-power lasers to accelerate the nanocraft: the radiation pressure of the laser beam on the light sail should accelerate the nanocraft~\cite{Marx,Redding}, which should quickly reach a velocity of some fractions of the speed of light. For example, the Breakthrough Starshot Initiative aims at developing a nanocraft able to travel at 1/5 of the speed of light in order to reach Alpha Centauri in about 20~years~\cite{2018AcAau.152..370P,breakthroughinitiatives}. Higher velocities can be reached at higher costs for the mission and there are no specific technical problems to reach 90\% of the speed of light with this approach.

In the case of a mission to study the closest black hole, we would need at least two nanocrafts, and with three or more nanocrafts we could get more precise measurements of the black hole and the physics around it. However, it may be more convenient to launch a single nanocraft, which could separate into two or more nanocrafts while it is near the target black hole. If the black hole is at 20-25~light-years and the nanocraft can travel at 1/3 of the speed of light, the nanocraft could reach the black hole in 60-75~years and we would need 20-25~more years to receive the results of the experiments, so the mission would last overall something like 80-100~years.


\section*{TESTS}

The tests that can be performed around the closest black hole depend on the actual instruments available on the nanocrafts. However, we can imagine at least three important tests: $i)$ test of the Kerr metric, $ii)$ test of the black hole event horizon, and $iii)$ test of possible variations of fundamental constants in a strong gravitational field (see Fig.~\ref{f-t}). In the case of a mission with two nanocrafts, ideally one of them (nanocraft~A) should orbit at a relatively large distance from the black hole and the other nanocraft (nanocraft~B) should orbit close to the black hole.

\subsection*{Kerr metric}

\begin{figure*}[t]
\centering
\includegraphics[width=0.27\linewidth]{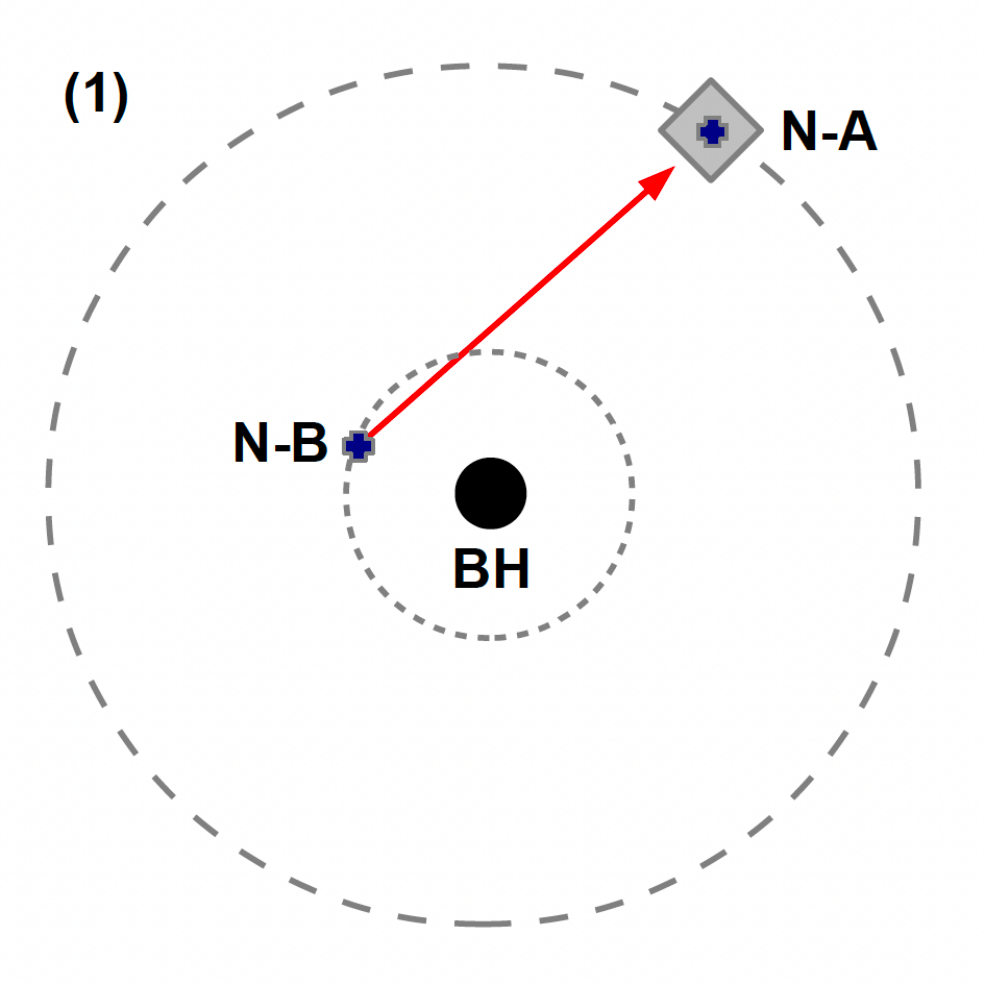}
\hspace{1.0cm}
\includegraphics[width=0.27\linewidth]{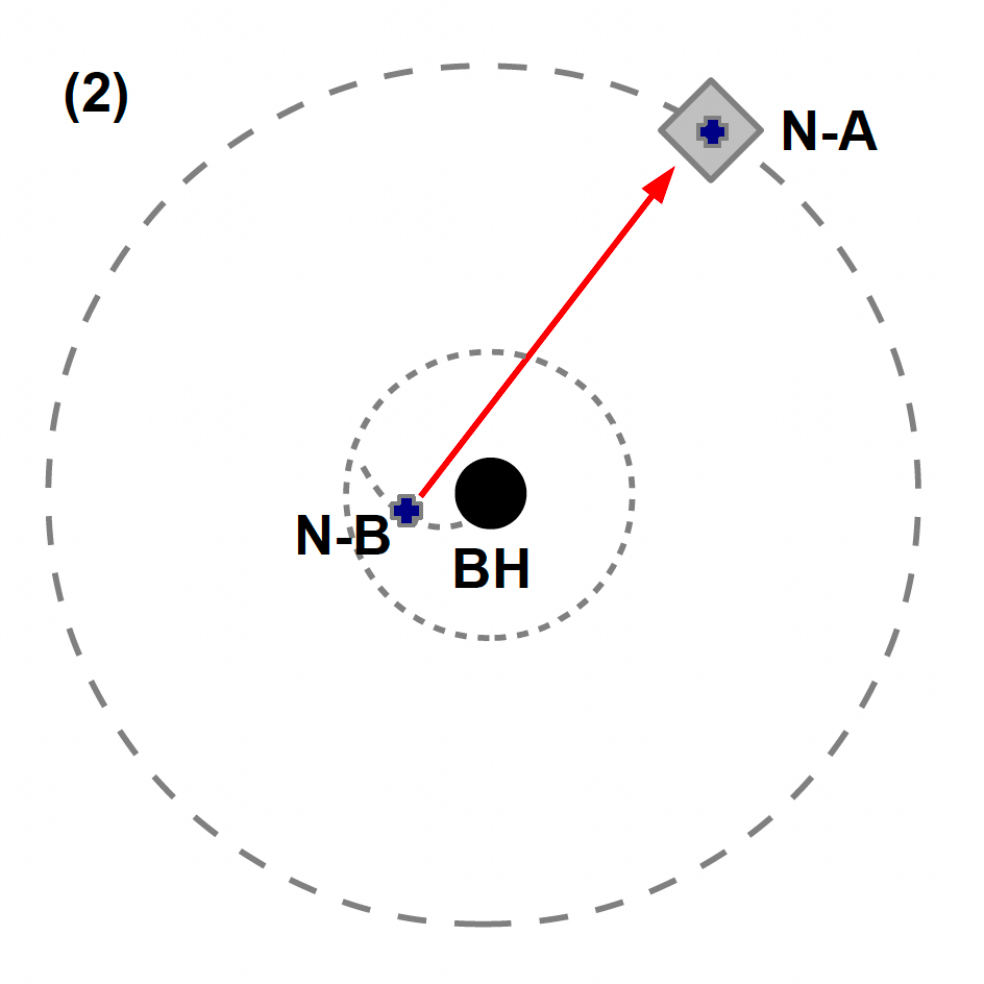}
\hspace{1.0cm}
\includegraphics[width=0.27\linewidth]{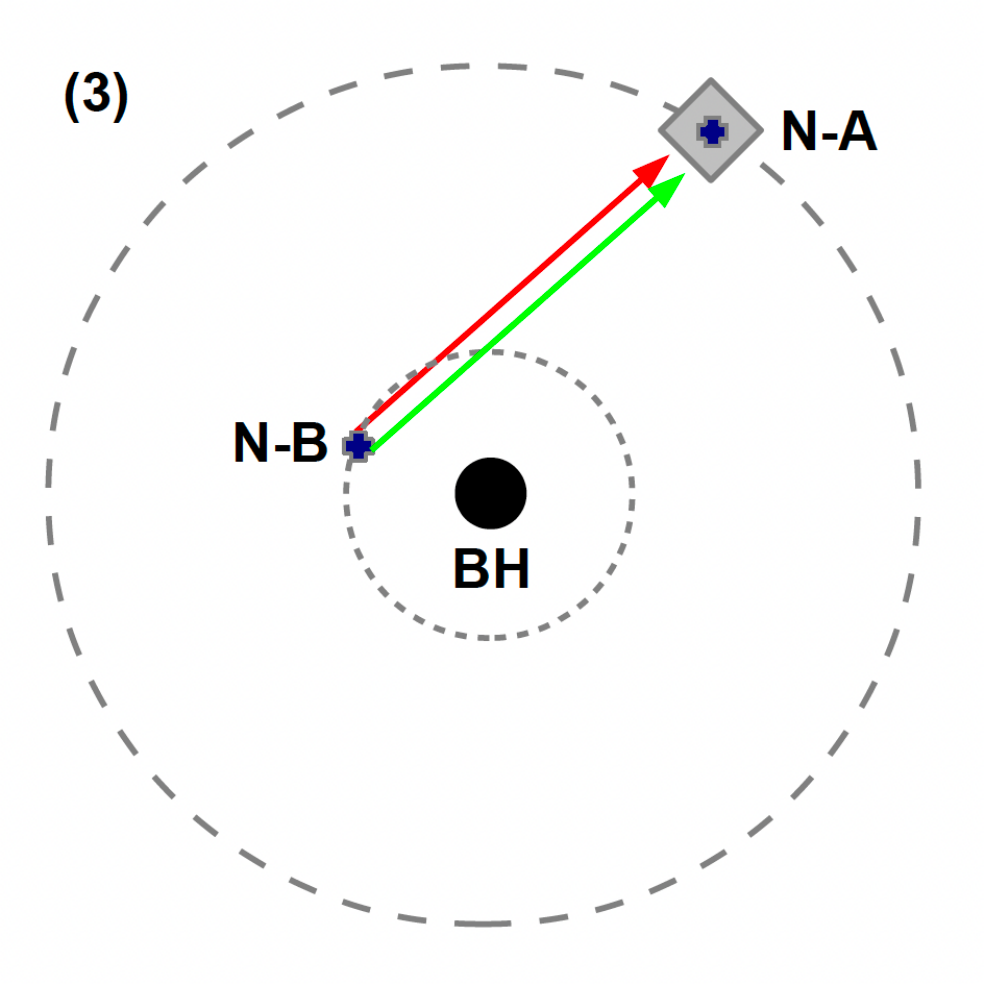}
\caption{{\bf Three possible important experiments around a black hole.} In these cartoons, the black hole (BH) is indicated by the central black circle, nanocraft~A (N-A) is indicated by the blue cross (the wafer of nanocraft~A) and the gray diamond (the light sail of nanocraft~A), and nanocraft~B (N-B) is indicated by the blue cross (the wafer of nanocraft~B). The orbits of nanocraft~A and nanocraft~B are indicated by the dashed curves. $(1)$ Test of the Kerr metric: nanocraft~A monitors the electromagnetic signal emitted by nanocraft~B and compares it with the theoretical predictions expected in the Kerr spacetime. $(2)$ Test of the event horizon: nanocraft~A monitors the electromagnetic signal emitted by nanocraft~B while nanocraft~B is falling onto the black hole. $(3)$ Test of possible variation of fundamental constants: nanocraft~A monitors two electromagnetic signals emitted by nanocraft~B to check whether they are affected by the same redshift.
\label{f-t}}
\end{figure*}

With a mission of two nanocrafts, nanocraft~A can observe nanocraft~B orbiting the black hole. Nanocraft~B can emit a stable electromagnetic signal, which is received by nanocraft~A (left panel in Fig.~\ref{f-t}). If the spacetime metric around the black hole is described by the Kerr solution, we can predict accurately the temporal evolution of the signal detected by nanocraft~A and we can then compare the theoretical predictions with the detected signal. If nanocraft~A can observe nanocraft~B for many cycles, the signal to noise ratio can be very high and we can potentially perform stringent tests of the Kerr metric.

If the spacetime metric around the black hole is not described by the Kerr solution, we can try to measure the mass and current moments of the spacetime~\cite{Hansen:1974zz}. Nanocraft~B should orbit relatively far from the black hole, where the multipole moment expansion works. The lowest order moments (mass, spin angular momentum, and mass quadrupole moment) can be measured earlier, while the measurement of higher and higher order moments requires longer and longer observational times. Moving nanocraft~B to orbits closer to the central object after measuring the lowest order moments can help to measure higher order moments. If the mission has more nanocrafts, the measurements of the multipole moments can be more precise and/or require shorter observational times.

\subsection*{Event horizon}

Roughly speaking, a black hole is a region of the spacetime in which gravity is so strong that nothing, nor even light, can escape to the exterior region. The event horizon is the boundary separating the black hole from the exterior region. In General Relativity and in many other theories of gravity, the final product of the gravitational collapse is a black hole with an event horizon. However, there are even models beyond General Relativity in which the final product of the gravitational collapse is an horizonless compact object.

Observations show that the gravitational potential energy of the accreting material around black holes is lost (namely it is not converted into radiation or kinetic energy of outflows) and these results are normally interpreted as an evidence for the existence of event horizons in astrophysical black holes~\cite{Narayan:2008bv,Broderick:2009ph}. Indeed, in the case of a compact object with a solid surface, we should expect to see the radiation released by the accreting matter that hits the solid surface of the compact object. However, some scenarios of horizonless compact objects can explain current observations. An example is the fuzzball paradigm, where the final product of the gravitational collapse is a {\it fuzzball}, namely a bound state of strings/branes without event horizon; see, e.g., Mathur \& Mehta~(2024)~\cite{Mathur:2024ify} for a review on the fuzzball paradigm. The matter accreting onto a fuzzball is not lost behind an event horizon and is instead quickly converted into fuzzball degrees of freedom on the fuzzball surface.

With a mission of two nanocrafts, nanocraft~A can observe nanocraft~B falling onto the compact object (central panel in Fig.~\ref{f-t}). In the presence of an event horizon, the signal from nanocraft~B should be more and more redshifted (formally without disappearing, as an observer should never see a test-particle crossing the event horizon in a finite time, but, in practice, at some point the signal leaves the sensitivity band of the receiver on nanocraft~A). If the compact object is a Kerr black hole, we can make clear predictions on the temporal evolution of the signal emitted by nanocraft~B. If the compact object is a fuzzball, the temporal evolution of the signal should be different and presumably stop instantly when nanocraft~B is converted into fuzzball degrees of freedom.

\subsection*{Fundamental constants}

In General Relativity and in any metric theory of gravity, atomic physics near a black hole is the same as atomic physics in our laboratories on Earth. This is a consequence of the fact that, in these theories, gravity universally couples to matter and therefore in any locally inertial reference frame the non-gravitational laws of physics reduce to those of Special Relativity~\cite{Will:2014kxa}. When gravity does not universally couple to matter, we can have phenomena like the variation of ``fundamental constants''~\cite{Bambi:2022lhq}. For example, there are models in which some of our fundamental constants (e.g., the fine structure constant or the electron mass) are not the true fundamental constants of the theory and depend on the vacuum expectation value of a scalar field, which may change in different gravitational fields; see, e.g., Davis et al.~(2016)~\cite{Davis:2016avf} for a toy model. While there are many studies reported in the literature on possible {\it temporal} variations of fundamental constants~\cite{Uzan:2002vq}, the case of variations in strong gravitational fields is not very explored and there are no observational constraints for gravitational fields stronger than those on the surface of white dwarfs~\cite{Berengut:2013dta}.

As an example, let us assume we want to test the possible variation of the fine structure constant $\alpha$ in the strong gravitational field of a black hole. We can consider two atomic transitions whose photon energies depend on $\alpha$ in a different way; for instance, the photon energy of a transition is proportional to $\alpha^2$ and the photon energy of the other transition is proportional to $\alpha^4$. With a mission with two nanocrafts, nanocraft~B can emit photons from the two atomic transitions and these photons can be detected by nanocraft~A (right panel in Fig.~\ref{f-t}). The photon energies at the detection point are different from the photon energies at the emission point because of gravitational redshift and Doppler boosting, but the shift of the two energies must be the same in a metric theory of gravity. From the detection of the photons of one of the two transitions, we can determine the photon redshift. From the detection of the photons of the second atomic transition, we can constrain/detect a possible variation of the fine structure constant $\alpha$.


\begin{figure*}[t]
\centering
\includegraphics[width=0.9\linewidth]{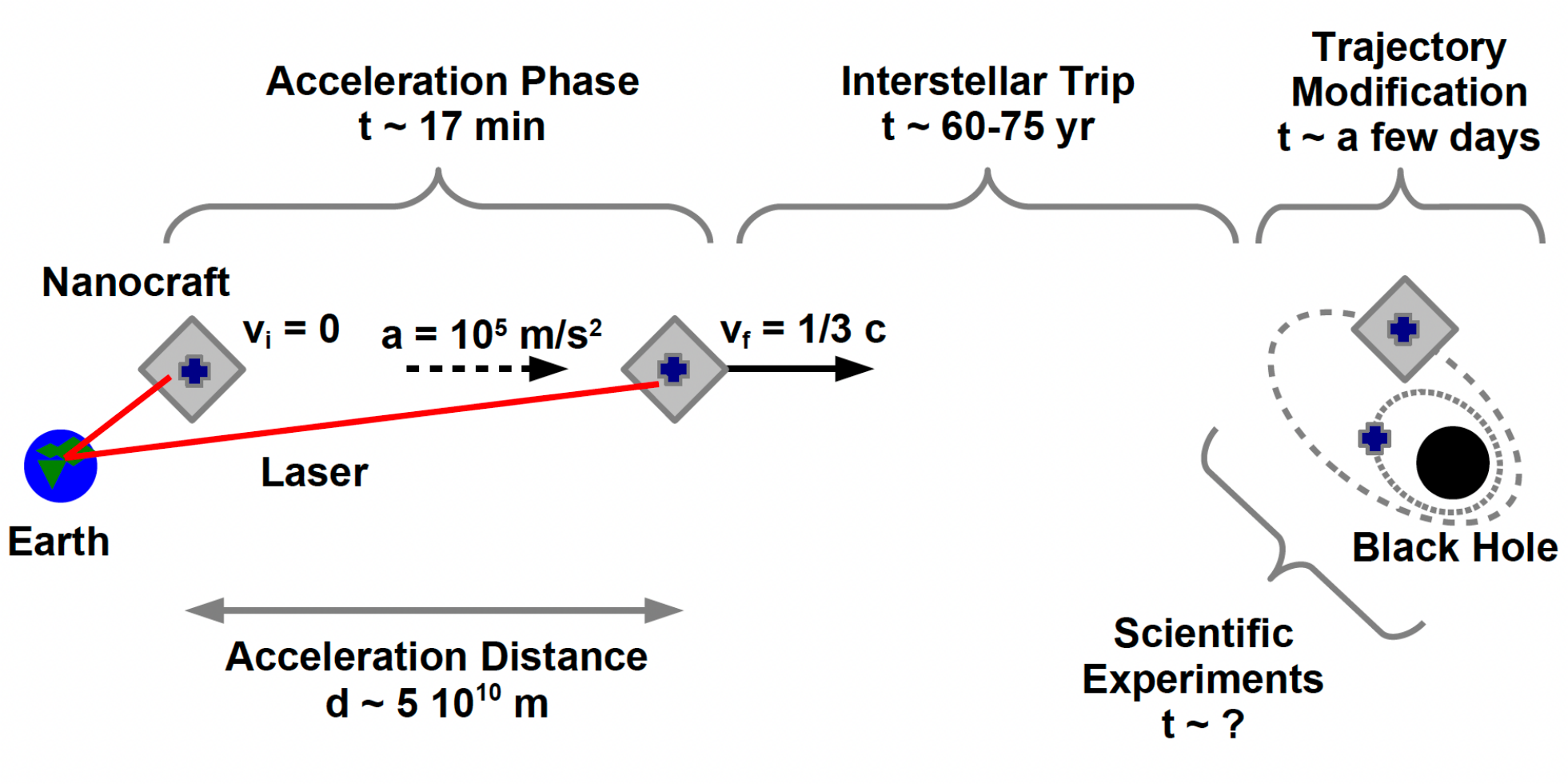}
\vspace{-0.2cm}
\caption{{\bf Phases of an hypothetical interstellar mission to the closest black hole.} Phase~1 (acceleration of the nanocraft): assuming that the target velocity is 1/3 of the speed of light and the nanocraft can endure a maximum acceleration of $10^5$~m~s$^{-2}$, the acceleration phase lasts around 17~minutes and the acceleration distance is $5 \cdot 10^{10}$~m. Phase~2 (interstellar trip): if the black hole is at 20-25~light-years from Earth and the nanocraft travels at 1/3 of the speed of light, the nanocraft can reach the black hole in 60-75~years. Phase~3 (approaching to the black hole and preparation for the scientific experiments): when the nanocraft is near the black hole, it should change its trajectory, get closer to the black hole, and move to the orbit assigned to start the scientific experiments; at the end of this phase, the nanocraft can separate into nanocraft~A, which should orbit relatively far from the black hole, and nanocraft~B, which should move closer to the black hole. Phase~4 (scientific experiments): the nanocrafts orbit around the black hole to perform all the planned scientific experiments and then send all collected data to Earth. 
\label{f-m}}
\end{figure*}

\section*{PHASES OF THE MISSION}

Fig.~\ref{f-m} shows the phases of such an hypothetical mission. Phase~1 is the acceleration of the nanocraft. From past studies, we can expect that the mass of the wafer is $m_w \sim 1$~g, the mass of the sail is $m_s \sim 1$~g, the area of the sail is $A_s \sim 10$~m$^2$, and the maximum acceleration that the nanocraft can endure is $a \sim 10^5$~m~s$^{-2}$~\cite{Kuhlmey25}. If we want the nanocraft to reach 1/3 of the speed of light, from simple Newtonian dynamics it follows that the acceleration phase should last $10^3$~s (around 17~minutes). In this case, the acceleration distance turns out to be $d \sim 5 \cdot 10^{10}$~m (which is roughly 1/3 of the distance Sun-Earth). Once the nanocraft reaches the target velocity, the laser is turned off, and Phase~1 ends.

Phase~2 is the interstellar trip: if the target black hole is at 20-25~light-years from Earth and the nanocraft travels at 1/3 of the speed of light, the nanocraft can reach the compact object in 60-75~years.

In Phase~3, the nanocraft should change its trajectory to approach the black hole and start orbiting around the compact object. This seems to be one of the most critical parts of the mission, because the nanocraft has to transfer from an unbound orbit to a bound orbit. When the nanocraft orbits the black hole, it can separate into nanocraft~A and nanocraft~B and nanocraft~B should move to an orbit closer to the compact object.

In Phase~4, the nanocrafts orbit around the black hole and perform all the scientific experiments in their program.

\section*{REQUIREMENTS AND CHALLENGES}

Today we do not know any black hole within 20-25~light-years of Earth and we do not have the technology to send a nanocraft to such an hypothetical black hole and test General Relativity. There are many challenges to solve in order to make a similar mission possible and each of them requires detailed studies. We can group these challenges into three wide categories: $i)$ the development of a nanocraft for interstellar missions, which is a goal common to other proposals devoted to study nearby stars and exoplanets, $ii)$ how to localize and approach the target black hole, and $iii)$ the development of the instruments to perform scientific experiments in the strong gravitational field of the black hole.

\subsection*{Nanocraft}

The development of a nanocraft for interstellar missions is certainly very challenging, but there are not very specific requirements for the case of an interstellar mission to test black holes. A recent review on the topic is Lin et al.~(2025)~\cite{Kuhlmey25}. The main challenges are the development of a laser array to accelerate the nanocraft, the development of suitable light sails, and the capability of the nanocraft to transfer data to Earth.

The laser to accelerate the nanocraft is a very critical part for every interstellar mission with nanocrafts. The first issue is the cost~\cite{2018AcAau.152..370P}. With current technology, the cost of the laser array would be around one trillion EUR, which is definitively beyond the budget of any scientific experiment. Since the trend is that the price per coherent watt halfs every four years, the cost of the laser array could be around one billion EUR within 30~years, which would be comparable to the budget of today's large space missions. A configuration using today's technology is discussed in Bandutunga et al.~(2021)~\cite{Bandutunga} and involves an array of $10^8$~lasers.

The sail membrane is another critical aspect for every mission. The sail should be lightweight, highly reflective, capable of surviving the laser irradiation and staying centered in the laser beam during the acceleration phase (while the laser beam will be rotated to compensate for the Earth's rotation). Meeting all these requirements simultaneously is extremely challenging, but may be achieved with optimized photonic designs~\cite{Kuhlmey25}.

Another issue common to all nanocraft missions is how to transfer the collected scientific data to Earth. The $10$~m$^2$ sail can serve as an antenna for downlink communication to Earth. In the case of two nanocrafts, nanocraft~A orbiting relatively far from the black hole can use the sail to transfer data to Earth. Nanocraft~B, orbiting closer to the black hole, does not need any sail to communicate with nanocraft~A, which is not far from nanocraft~B. At the same time, if nanocraft~B had a sail it could not move very close to the black hole, as the tidal forces near the black hole could easily destroy its sail.

\subsection*{Localization and approaching}

A precise and accurate localization of the black hole is another critical part of the mission. First, it is necessary to get a precise and accurate localization of the black hole in the sky before the launch of the nanocraft. This allows us to send the nanocraft as close as possible to its destination. Second, when the nanocraft is ``near'' the black hole, it should be able to modify its trajectory to get closer to the black hole and get ready for the scientific experiments.

The localization of the black hole in the sky depends on the technique(s) used to discover the black hole and determine its location in the sky with the best possible precision and accuracy~\cite{loc}. 

In the case of the method proposed in Jana et al.~(2025)~\cite{Jana:2024hks} based on the detection of gravitational waves, with current detectors and algorithms it is difficult to get a precise and accurate localization of the source. Current approaches to determine the position of gravitational wave events in the sky are based on the delay of the gravitational wave signal among different detectors: at least three detectors are required and the localization in the sky can be significantly improved with a higher number of detectors~\cite{Klimenko:2011hz}. On the other hand, the future gravitational wave observatory Einstein Telescope~\cite{Punturo:2010zz} will have a triangular geometry and will be able to determine the location of gravitational wave transients in the sky even without the detection of other observatories. In general, the accuracy of the position of the source in the sky depends on the gravitational wave amplitude, the duration of the signal, the algorithm used, the accuracy of the theoretical predictions of the waveforms, etc. As of now, there are no specific studies on the localization of nearby black holes emitting gravitational waves as discussed in Jana et al.~(2025)~\cite{Jana:2024hks}. Since we should be able to observe the source many times (every time it is illuminated by a sufficiently powerful photon flux), we could improve its sky location over time.

Assuming to be able to localize the source in the sky with an accuracy of 1~deg$^2$ and that the source is at 20~light-years from us, we could direct the nanocraft to the location of the black hole during the acceleration phase (Phase~1) with a precision of about 22k~AU, which is certainly not enough for the nanocraft to find the black hole. We may need to localize the black hole in the sky with an accuracy of 1~arcmin$^2$ (which is currently unfeasible for a gravitational wave event) and in such a case the nanocraft could approach the black hole at about 370~AU. At such a distance, the gravitation force of a 10~$M_\odot$ black hole on the nanocraft would be comparable to the gravitation force of the Sun on the nanocraft at the orbit of Neptune.

In the case the black hole is detected with electromagnetic techniques, such as the method proposed in Murchikova \& Sahu (2025)~\cite{Murchikova:2025oio}, the localization of the black hole in the sky can be very precise and accurate and therefore it does not represent a critical requirement for the mission. 

During the interstellar trip (Phase~2), the trajectory of the nanocraft is that resulting from the acceleration phase (Phase~1). When the nanocraft is sufficiently close to black hole, it should modify its trajectory to move to the orbit assigned to perform the scientific experiments. Since the black hole does not emit radiation, the nanocraft may identify the black hole with some onboard instrument capable of detecting the gravitational field of the compact object. In the case of a mission to a star, the nanocraft can know the location of the source by detecting its electromagnetic radiation, which is not possible in the case of an isolated black hole. 

The most challenging phase of the mission may be related to how the nanocraft can transfer from an unbound to a bound orbit and start orbiting around the compact object. All possible solutions should be considered carefully. In the case the transfer were not possible, we may redesign the mission to perform the scientific tests when the nanocraft is passing close to the black hole. For example, when the nanocraft is close to the black hole it may separate into a mother-nanocraft (with wafer and sail) and a number of small nanocrafts (without sails). The nanocrafts could communicate with each other by exchanging electromagnetic signals. The mother-nanocraft could compare the trajectories of the small nanocrafts to those expected in a Kerr spacetime and send the data to Earth.

\subsection*{Scientific experiments}

Spacecrafts orbiting around an isolated black hole could {\it potentially} perform very precise tests of General Relativity, as there are no complications related to the astrophysical environment like in the case of an accreting black hole. On the other hand, the challenges are related to the fact that we can only send very small nanocrafts, so all scientific instruments should be in a 1~g wafer, and the nanocraft has to send all data to Earth, which may be at 20-25~light-years from the black hole.

We need at least two nanocrafts to test the nature of the black hole, and with three or more nanocrafts we can expect more precise and accurate measurements. However, it is likely more convenient to launch a single nanocraft with a wafer and a light sail. When the nanocraft reaches the black hole, it may separate into two or more parts. In the case of the two nanocrafts discussed before, nanocraft~A would orbit relatively far from the black hole: it should monitor the signal emitted by nanocraft~B and send the data to Earth. Nanocraft~A could use the light sail as an antenna to transfer the data to Earth. Nanocraft~B should orbit closer to the black hole: it does not need a light sail to communicate with nanocraft~A because the two nanocrafts are not too far each other and, at the same time, a light sail may not survive close to the black hole because of tidal forces.

\subsubsection*{Kerr metric} 

If we want to test the Kerr geometry around the black hole, it is convenient to have nanocraft~B orbiting as close as possible to the compact object, where the gravitational field is stronger and small deviations may be easier to detect. On the other hand, if we knew that the spacetime metric around the compact object is not described by the Kerr solution, it may be more interesting to measure the multipole moments of the spacetime in order to reconstruct the actual spacetime geometry: in such a case, nanocraft~B should orbit at a certain distance from the compact object, where the multipole moment expansions is valid. After measuring the black hole mass, spin, and quadrupole moment, nanocraft~B could move to orbits closer to the compact object to infer higher order moments.

Solar System experiments can test General Relativity at the first post-Newtonian order~\cite{EventHorizonTelescope:2020qrl} with a precision at the level of $10^{-5}$~\cite{Will:2014kxa} and cannot constrain corrections beyond the first post-Newtonian order. X-ray observations currently provide the most stringent test of the Kerr metric and can constrain corrections beyond the first post-Newtonian order at the level of 10\%~\cite{Tripathi:2020yts}. The European space-based gravitational wave observatory LISA, which is currently scheduled to be launched in 2035, will be able to test the nature of black holes by detecting EMRIs (extreme mass ratio inspirals, in which a stellar-mass compact object like a stellar-mass black hole, neutron star, or white dwarf orbits around a supermassive black hole). Since LISA can observe millions of orbits of a stellar-mass compact object around a supermassive black hole, we can expect to be able to constrain corrections of order $(R_{\rm g}/R)^3$ in the Kerr metric, where $R_{\rm g} = G_{\rm N}M/c^2$ is the gravitational radius of the black hole and $R$ is the orbital radius, at the level of 1\% and even better, maybe up to a level of $10^{-4}$ for some systems~\cite{Barack:2006pq}.

In the case of the experiment sketched in the left panel in Fig.~\ref{f-t}, nanocraft~A can monitor the orbital frequency of nanocraft~B and the redshift of its signal. Since frequency precision increases by observing many cycles, monitoring the orbital frequency of nanocraft~B can provide more stringent constraints on possible deviations from the Kerr geometry. If we want to test corrections of order $(R_{\rm g}/R)^3$ in the Kerr metric, we have to constrain corrections of order $(R_{\rm g}/R)^3$ in the orbital frequency. Assuming that the black hole mass is 10~$M_\odot$, the orbital frequency at the innermost stable circular orbit (ISCO) of a non-rotating black hole of General Relativity is 220~Hz; see Table~\ref{tab} and Bambi~(2017)~\cite{Bambi:2017khi} for the derivation. If nanocraft~B orbits at the ISCO radius and nanocraft~A monitors the signal emitted by nanocraft~B for about 3~days, we can constrain corrections of order $(R_{\rm g}/R)^3$ in the Kerr metric at the level of $10^{-6}$. If nanocraft~A monitors the signal from nanocraft~B for about 1~month, we can constrain corrections of order $(R_{\rm g}/R)^3$ at the level of $10^{-7}$. For a 10~month observation, we can constrain those corrections at the level of $10^{-8}$. The key-point here is that an isolated black hole is a very clean system, without environmental effects that can spoil our measurements: the system is a black hole in vacuum and the tiny nanocrafts orbiting the black hole are completely negligible.

\begin{table}[t]
\centering
\renewcommand\arraystretch{1.5}
\caption{{\bf Orbital frequency and tidal acceleration.} Orbital frequency and Newtonian tidal acceleration on a 1~cm nanocraft around a 10~$M_\odot$ non-rotating black hole of General Relativity at different orbital radii. In the first row, it is shown the tidal acceleration at the black hole event horizon $R_{\rm H}$, while there is no orbital frequency because there are no circular orbits~\cite{Bambi:2017khi}. In the second row, it is shown the orbital frequency and tidal acceleration at the innermost stable circular orbit (ISCO) $R_{\rm ISCO}$. In the third and fourth rows, the orbital radius refers to the radial coordinate in Schwarzschild coordinates. In the case of a nanocraft with a 1~m light sail, the tidal acceleration on the light sail can be up to 100~times that on a 1~cm nanocraft and therefore a light sail cannot survive at $R_{\rm ISCO}$ or at 10~$R_{\rm g}$.}
\vspace{0.15cm}
\begin{tabular}{|c|c|c|}
\hline\hline
\hspace{0cm} Orbital Radius \hspace{0cm} & \hspace{0cm} Orbital Frequency \hspace{0cm} & \hspace{0cm} Tidal Acceleration \hspace{0cm} \\
 & (Hz) & (m~s$^{-2}$) \\
\hline\hline
$R_{\rm H}$ & -- & $5 \cdot 10^5$ \\
$R_{\rm ISCO}$ & 220 & $2 \cdot 10^4$ \\
10~$R_{\rm g}$ & 102 & $4 \cdot 10^3$ \\
100~$R_{\rm g}$ & 3.23 & $4$ \\
\hline\hline
\end{tabular}
\label{tab}
\end{table}

\subsubsection*{Event horizon}

There are potentially two critical points to test the black hole event horizon: $i)$ nanocraft~B should endure the tidal forces produced by the gravitational field of the black hole, and $ii)$ nanocraft~A should be able to detect even extremely redshifted signals from nanocraft~B.

Actually point~$i)$ is not a particularly critical issue, as nanocraft~B is small. Table~\ref{tab} shows that the tidal acceleration of a 10~$M_\odot$ black hole on a 1~cm nanocraft at the event horizon is comparable to the acceleration experience by the nanocraft in Phase~1. While Table~\ref{tab} reports the Newtonian tidal acceleration, it is enough for a rough estimate of the acceleration that the nanocraft should be able to endure to avoid its destruction.

The theoretical prediction of General Relativity is that nanocraft~A should observe nanocraft~B falling onto the black hole and approaching the event horizon forever, without reaching it. In reality, the signal emitted from nanocraft~B should become more and more redshifted, so at some point nanocraft~A would be unable to observe nanocraft~B any longer either because the frequency of the signal is too low to be detected by the instruments onboard nanocraft~A or because the signal is too weak. The challenge here is to have instruments onboard nanocraft~A that could observe nanocraft~B approaching the black hole event horizon for as long as possible, as in the absence of an event horizon this would increase the chances to observe new physics. Unfortunately, models beyond General Relativity such as the fuzzball scenario from string theories do not make clear predictions until now. We can only try to extend the observation of nanocraft~B as close as possible to the black hole, but there are no clear targets to falsify the fuzzball paradigm or other models. As of now, there are no quantitative model-independent constraints on the existence of event horizons from current astrophysical observations, so a comparison between the constraining powers of the nanocrafts and astrophysical observations is not possible.

\subsubsection*{Fundamental constants}

Even in the case of tests of the variation of fundamental constants, the challenge is that the instruments for the experiment should be adjusted in the nanocrafts. If we want to test possible variations of the fine structure constant $\alpha$, we can compare wavelengths of different atomic transitions. Non-relativistic spectra depends mainly on the Rydberg constant $R_\infty = \alpha^2 m_e c/2h$, where $m_e$ is the electron mass, $c$ is the speed of light, and $h$ is the Planck constant. The fine structure depends on $R_\infty \alpha^2$. In the case of fine-structure doublets, the frequency splitting between the two lines of the doublet is~\cite{bp77}
\be
\Delta\nu = \frac{\alpha^2 Z^4 R_\infty}{2 n^3} \, , \nonumber
\ee
where $Z$ is the proton number of the nucleus and $n$ is the principal quantum number. If $\nu$ is the mean frequency of the doublet, $\Delta\nu/\nu \sim \alpha^2 Z^2$ and is independent of the photon redshift, so a variation of $\Delta\nu/\nu$ can be attributed to a variation of $\alpha$.

Fine-structure doublets are used to constrain possible temporal variations of $\alpha$ with observations of spectra of distant astrophysical objects. In the case of an interstellar mission, the experiment would require that nanocraft~B sends the radiation of the doublet to nanocraft~A and nanocraft~A measures $\Delta\nu/\nu$ and compares its value to that in our laboratories on Earth. For example, the sodium D-lines are split by 515~GHz. If the instruments onboard nanocraft~A can measure $\Delta\nu/\nu$ with a precision of 1\%, we can test possible variations of $\alpha$ at the level 0.5\%.


\section*{ROADMAP}

The first step towards an interstellar mission to test black holes is to find a black hole within an acceptable distance of Earth. In this letter, I argued that the closest black hole may be within 20-25~light-years of Earth, but this is just a rough estimate. Apart its uncertainty, only the discovery of a nearby black hole can boost the efforts towards a real interstellar mission and permits us to evaluate the actual chances to send a spacecraft to test General Relativity in the strong gravitational field of the compact object. If the distance of the closest black hole is larger than 25~light-years and we want to remain with a mission under 100~years, we need that the nanocraft velocity exceeds 1/3 of the speed of light. While there are no specific technical problems to reach 90\% of the speed of light with nanocrafts, the total cost of the mission can increase significantly. If the black hole is at 45~light-years (13.8~pc) and the nanocraft velocity is 90\% of the speed of light, the nanocraft can reach the black hole in 50~years and we need 45~years to receive the results of the experiments on Earth.

After the discovery of a nearby black hole, we can work to get a precise and accurate location of the source in the sky. From the simple estimates presented in this letter, it seems we need to determine the location of the black hole at least with an accuracy of about 1~arcmin$^2$. At the same time, we can start thinking about the mission. As discussed before, there are parts of the interstellar mission that are common to other proposals of interstellar missions with nanocrafts and there are other parts that are specific for a mission to test a black hole. The proposal presented in this letter is certainly more challenging than other proposals to send nanocrafts to study nearby stars and exoplanets and therefore it will be unlikely the first mission of this kind. The technology of the laser to accelerate the nanocraft, of the sail material, and of the communication instruments to send data from a far away nanocraft to Earth will be likely tested with other interstellar missions. However, it will be also important to develop the technology specific for a mission to a black hole, which was discussed in the previous section. 

If there exists a black hole within 20-25~light-years of Earth and we find a way to discover it, it is probably only an issue of time to reach the technology necessary to send a probe to the object and obtain very valuable information about black holes and General Relativity that might be difficult to obtain in other ways.


\section*{CONCLUDING REMARKS}

In this letter, I argued that the possibility of an interstellar mission to study a black hole is not completely unrealistic even if it is certainly very speculative and extremely challenging. I presented this idea as a concept and there are clearly many issues to study and solve before starting designing a similar mission (e.g., how we can find the closest black hole, how we can determine its location in the sky with sufficient accuracy, how we can send our spacecrafts to the right place, all the technology of the spacecrafts and of their instruments has to be develop, etc.).

For a mission with a nanocraft traveling at 1/3 of the speed of light and for a black hole at a distance of 20-25~light-years, the whole mission may last 80-100~years: the nanocraft can reach the object in 60-75~years, study the black hole, and we can receive the data after 20-25~years. If the nanocraft can travel at a velocity close to the speed of light, the mission could last 40-50~years. These are very rough estimates, which strongly depend on the exact distance of the closest black hole (which is currently completely unknown) and the actual velocity of the nanocraft.

The possibility to explore the spacetime around a black hole with small spacecrafts promises to provide extremely valuable information about the nature of astrophysical black holes and the predictions of General Relativity in the strong field regime that may be impossible to obtain with astrophysical observations. The very long duration of a similar mission may discourage the efforts of the scientific community and the support of funding agencies and Governments. However, we should realize that most of the future experiments in particle physics and astrophysics will likely require long times (for preparation, construction, and data collection) and the work of a few generations of scientists, be very expensive, and in many cases we will not have other options if we want to make progress in a certain field.

\newpage

\section*{RESOURCE AVAILABILITY}

\subsection*{Lead contact}

Requests for further information and resources should be directed to and will be fulfilled by the lead contact, Cosimo Bambi (bambi@fudan.edu.cn).

\subsection*{Materials availability}

This study did not generate new unique materials.

\subsection*{Data and code availability}

This study did not analyze data and did not develop/use codes.

\section*{ACKNOWLEDGMENTS}

I thank Swarnim Shashank and Debtroy Das for useful discussions and feedback.
This work was supported by the National Natural Science Foundation of China (NSFC), Grant No.~12261131497 and No.~12250610185.

\section*{AUTHOR CONTRIBUTIONS}

Conceptualization, writing, C.B.

\section*{DECLARATION OF INTERESTS}

The author declares to be a member of the Advisory Board of iScience.


\begin{thebibliography}{99}
\bibitem{loc}
In principle, we may use a certain technique to discover the closest black hole and another technique to determine/improve its location in the sky.
\end{thebibliography}
\end{document}